\documentclass[twocolumn,aps,pre]{revtex4}
\usepackage{graphicx,color}
\begin{document}
\title{\bf{Fixed points and boundary layers in  asymmetric simple 
exclusion processes}}
\author{Sutapa Mukherji}
\affiliation{Department of Physics, Indian Institute of Technology, 
Kanpur 208016, India }
\date{\today}
\begin{abstract}
In this paper, we show how a fixed point based boundary layer analysis 
can be used to understand phases and phase transitions in 
asymmetric simple exclusion processes (ASEPs) 
with open boundaries. In order to 
illustrate this method, we choose a two-species ASEP which 
has interesting phase transitions not seen in the one-species case. 
We also apply this method to the single species problem where 
the analysis is  simple but nevertheless quite insightful.   
\end{abstract}
\maketitle

Asymmetric simple exclusion process (ASEP)\cite{ligett}
 in its simplest form involves 
  particles moving on a one-dimensional (1D) 
lattice with a bias in one 
specific direction. Due to the  mutual exclusion rule, 
the occupancy of a  site by more than one particle is ruled out. 
The biased motion of particles is responsible for a  particle    
current which is a typical indication of the system being away from the
thermodynamic equilibrium. With open boundaries, the system needs
 to be coupled to particle reservoirs 
which maintain  constant densities at the boundaries.  
All these non-equilibrium systems 
after sufficiently long time settle in a steady state where 
properties like the average density of  particles, current  
etc are independent of time.
A striking  feature of the open system is the boundary induced 
phase transition where the tuning parameters are the boundary 
densities \cite{krug}. Various bulk phases   characterized 
through distinct current  or particle-density profiles
are usually represented  in a phase diagram in the space of 
the  boundary densities.


Unlike equilibrium systems which carry no current, for ASEP, 
 the information  about the  
change in the  boundary condition (BC)  is 
mediated  up to the bulk by the particle current. 
Therefore,  the issue as how 
the BCs affect the  
 boundary layers  holds the key 
to understand the boundary induced phase transitions in ASEP.     
Recent studies \cite{smsmb,smb} have shown that 
the boundary layer analysis is a very  general approach to probe
different models of  ASEP. 
In this approach, starting with the steady 
state hydrodynamic equation, a
 uniform approximation for the 
density valid both in the bulk and in the boundary region is  
developed.
For particle non-conserving ASEP,
this method  shows that  the formation of a discontinuity 
in the bulk-density profile  
  (or a localized  shock) is due to a  
deconfinement of  a boundary layer from the boundary. 
This deconfinement is shown to be 
 associated with the divergence of the width 
of the boundary layer \cite{smsmb}.

In this paper, we show how a fixed point description of the 
boundary layers and the phase-plane trajectories connecting 
the fixed points
 can be used to map the bulk phase diagram and identify 
the phase transitions. As an example, this method is applied 
to a    particle number conserving ASEP  
that  involves two species of particles \cite{mukamel,zia}. 
In addition, we also 
consider the single species system, where the analysis is  
especially simple.

In the two species model considered in this paper,  particles,
denoted by A and B,   move in opposite directions 
on a 1D lattice  
of length $L$ with  $N$ lattice points  obeying the 
following hopping rules  
$A0\rightarrow 0A$ with  rate $1$,
$0B\rightarrow B0$ with rate $1$ and 
$AB\rightarrow BA$  with rate $2$.
Here $0$ denotes an empty lattice site. 
While A type particles are injected (withdrawn) at the 
left (right) boundary, 
the reverse happens for B type 
particles at these boundaries. We assume that 
these injection and withdrawal rates 
effectively correspond to boundary 
reservoirs with fixed particle densities which are 
the BCs for our problem. Two species models 
with specific boundary  rates are known for  
spontaneous symmetry breaking  among the two species 
even when their dynamics is completely symmetric. \cite{comment1}. 
While systems specified through boundary rates show spontaneous 
symmetry breaking, 
this is  not seen in systems specified through boundary reservoir 
densities \cite{popkov}.
 Our analysis is valid for arbitrary  
boundary  densities.
Domain-wall based studies  with
boundary densities \cite{popkov} predict  two 
first order transitions  taking the system from a low to a high density 
jammed phase via an intermediate phase which is  not seen in single
species systems.
In the steady state, the particle number 
conservation leads to a flat density profile in the bulk.  
Domain wall   arguments show that unlike 
the  high or low density phase where, 
as in the single species case, 
 the bulk densities are same as one of the boundary densities,
the intermediate phase  has bulk-density not directly related 
to the boundary densities. 
Thus, it is important understand what  basic mechanism 
can be  responsible for the existence of an intermediate  phase and  
 how the value of the bulk density 
in this   phase is selected. 
The fixed point description presented here  unveils these issues.  
 
The hydrodynamic formulation begins with the  statistical 
averaging  of the master equation that describes the time 
evolution of the probabilities of a given site being 
 occupied by either one of the two types of particles. A continuum limit, 
as the lattice spacing, $a\rightarrow 0$ and $N\rightarrow \infty$ with 
$L=Na$ remaining finite, is, then, taken for the resulting equation. 
The differential equations, thus obtained, involve the 
average particle-densities 
$\rho(x,t)$ and $\eta(x,t)$ for $A$ and $B$ type particles with 
$x$ and $t$ as continuous variables representing position and time 
and the 
corresponding current-densities $j_\rho$ and $j_\eta$ as functions 
of $\rho$ and $\eta$.
In the first order  in $\epsilon=a/2$, 
the continuum equations are
\begin{eqnarray}
\frac{\partial \rho}{\partial t}+\frac{\partial j_\rho}{\partial x}=
\epsilon \frac{\partial}{\partial x}
((1+\eta)\frac{\partial \rho}{\partial x}-\rho 
\frac{\partial \eta}{\partial x})\label{hydro1}\\
\frac{\partial \eta}{\partial t}+\frac{\partial j_\eta}{\partial x}=
\epsilon \frac{\partial}{\partial x}
((-\eta)\frac{\partial \rho}{\partial x}+(1+\rho) 
\frac{\partial \eta}{\partial x}).\label{hydro2}
\end{eqnarray}
The stationary state of the model 
 on a ring is given by a product 
measure and the exact stationary fluxes, $j_\rho$ and $j_\eta$, are 
$j_\rho=\rho(1-\rho+\eta)\ \ {\rm and}\ \ 
j_\eta=-\eta(1-\eta+\rho)$.

The diffusion like terms in equations (\ref{hydro1}) and (\ref{hydro2})
 are small  in the $\epsilon \rightarrow 0$ 
limit. However,  these terms are  crucial for generating uniform 
approximations of the solutions. The stationary bulk-densities 
($\rho_b$ and $\eta_b$) of constant values, 
satisfy the differential equations trivially.
These constant solutions cannot satisfy the BCs at  
both the ends. In order to satisfy the BCs, 
 boundary layers are formed over  narrow regions near 
one or both the boundaries. 
 To focus
on this narrow region, we need to scale the position  variable $x$ 
as $\tilde x=(x-x_0)/\epsilon$, where $x_0$ 
specifies the location of the boundary layer,  
and express the stationary equations 
in terms of this new variable. Here, $x_0=0$$(1)$ 
for left(right) boundary layer which merges to the bulk 
particle-density profile  asymptotically as 
$\tilde x \rightarrow + \infty(-\infty)$.  
 A little algebraic manipulation and 
integrations of the stationary versions of  equations 
(\ref{hydro1}) and (\ref{hydro2}) lead to     
\begin{eqnarray}
p'(\tilde x)=c+q(1-p), \ {\rm and}\nonumber\\ 
(1+p)\ q'(\tilde x)=d+p+q(c-p q),\label{linearcomb}
\end{eqnarray}
where $p=\rho+\eta$ and $q=\rho-\eta$ with values lying  
within the range $[0,1]$ and $[-1,1]$, respectively. Here,
 prime denotes a derivative with respect to $\tilde x$
and $c$ and $d$ are the integration constants.
The saturation of the boundary layers located, for example, 
  near $x=L$, 
to bulk  as $\tilde x\rightarrow -\infty$, requires 
$c=-q_b(1-p_b)$  and  
$d=q_b^2-p_b$,
where $p_b$ and $q_b$ are the bulk values of $p$ and $q$, respectively. 
Since the bulk profile is flat,  same $c$ and $d$ are applicable  
in case of boundary layers appearing near both the boundaries. 
 For particle 
non-conserving systems, where the  bulk profile is $x$ dependent, 
$p_b$ and $q_b$ are the  values of $p$ and $q$ at the point where the  
boundary layer merges the bulk. Although 
this approach is applicable to  particle non-conserving systems, 
we restrict to the conserved 
case to avoid algebraic complications.  Because 
of same $c$ and $d$ and the translational invariance, 
the same phase portrait is applicable to  both the  boundaries.

\begin{figure}[htbp]
   \includegraphics[width=3in,clip]{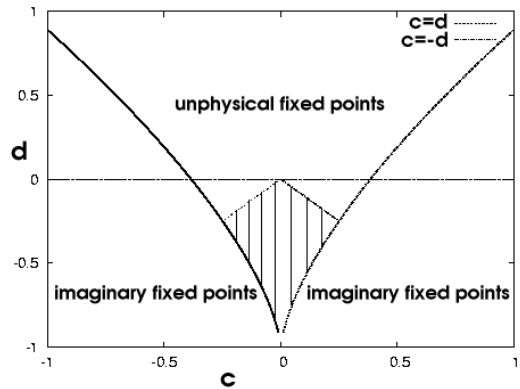}
    \caption{Nature of the fixed points in the physically acceptable 
range of the c-d plane. On the bold solid line and the dashed line 
the discriminant of equation (\ref{fp2}) is zero. Fixed points are 
imaginary in the region 
outside these two lines. In the striped region enclosed by these lines 
and the lines 
$d=c$ and $d=-c$, there are physically acceptable fixed points. 
In the rest of the space, not a single fixed point in the set of 
three has any physically acceptable value.}
\label{fig:c-d-plane}
\end{figure}

To understand the origin of  the three phases, it is useful to  
study the stability properties of the fixed points of equations in 
 (\ref{linearcomb}).  A combined 
 single  first order differential 
equation involving $\frac{dp}{dq}$ leads to the solution, $p(q)$,
 as a function of $q$, 
for a given BC at $x=1 \ {\rm or} \ 0$. These solutions 
are the phase-plane trajectories flowing to the appropriate fixed points.
  The boundary layers are these trajectories 
seen as functions of $\tilde x$.
Since these flows approach the 
fixed points in $\tilde x\rightarrow\pm\infty$, 
the bulk-densities correspond to one fixed point. 
Also, by the choice of the constants $c$ and $d$, the 
equations  are guaranteed to have a fixed point that corresponds 
to the bulk-densities. 
There exists three  fixed points, $(p^*,q^*)$, which 
are the solutions of the equations 
\begin{eqnarray}
p^*=(c+q^*)/q^*\ \ \ {\rm and}\label{fp1}\ \ \  
(q^*)^3-(d+1)q^*-c=0. \label{fp2}
\end{eqnarray}
Although the physically acceptable range of both 
$c$ and $d$ is   $[-1,1]$, there is only a narrow window 
in this $c-d$ space (see the striped region in 
figure \ref{fig:c-d-plane}),  where the fixed
 points can represent a meaningful density profile (within the 
physical domain {\it i.e.} the values of $p$ and $q$ lie in the range 
$[0,1]$ and $[-1,1]$ respectively and $0\le \rho,\eta \le 1$ ).
 Outside this 
window, the fixed points are either imaginary, or do not have a single 
 member that can represent the bulk. A stability analysis shows  that 
out of the three fixed points, one is unstable, one is saddle and one 
is stable. 
Based on their locations on the $p-q$ plane, we call the stable and 
unstable fixed points as right (R) or left (L). The saddle fixed 
point is always denoted as S.

\begin{figure}[htbp]
   \includegraphics[width=3in,clip]{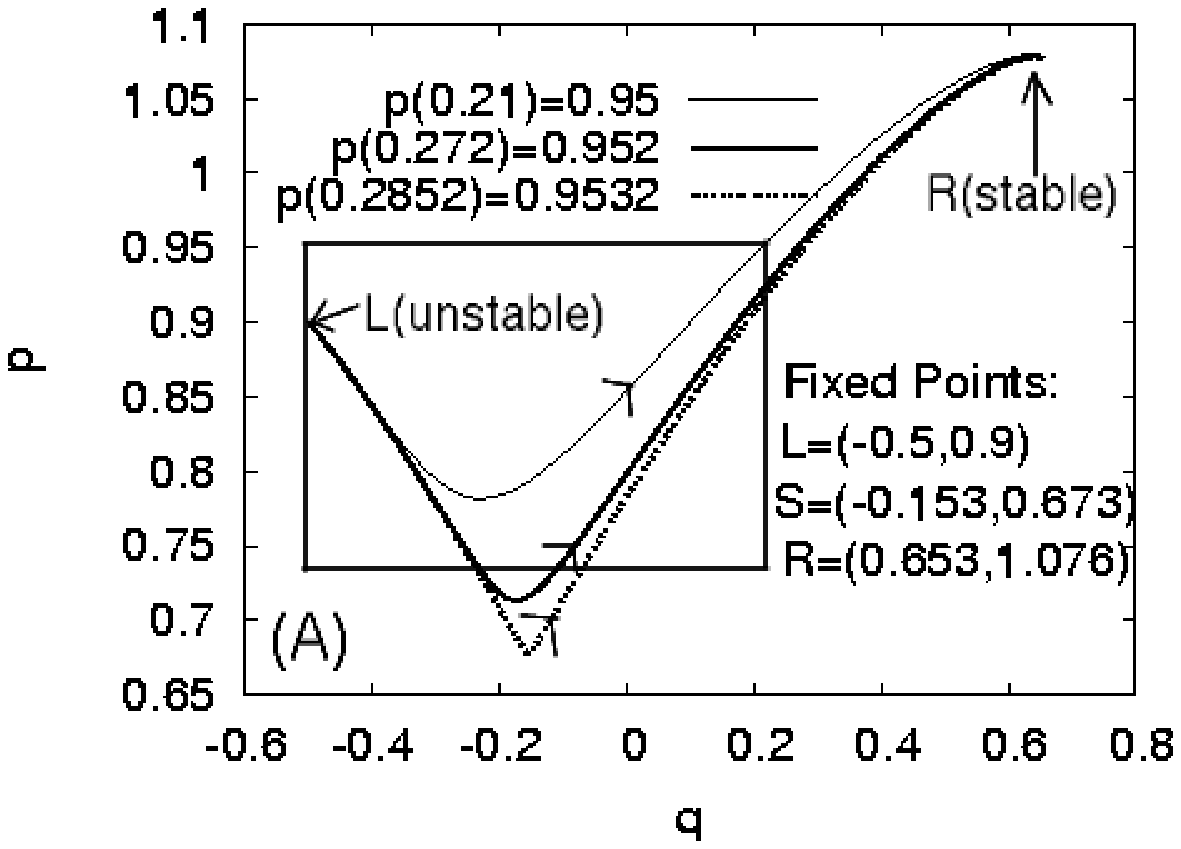}
\includegraphics[width=2.7in,clip]{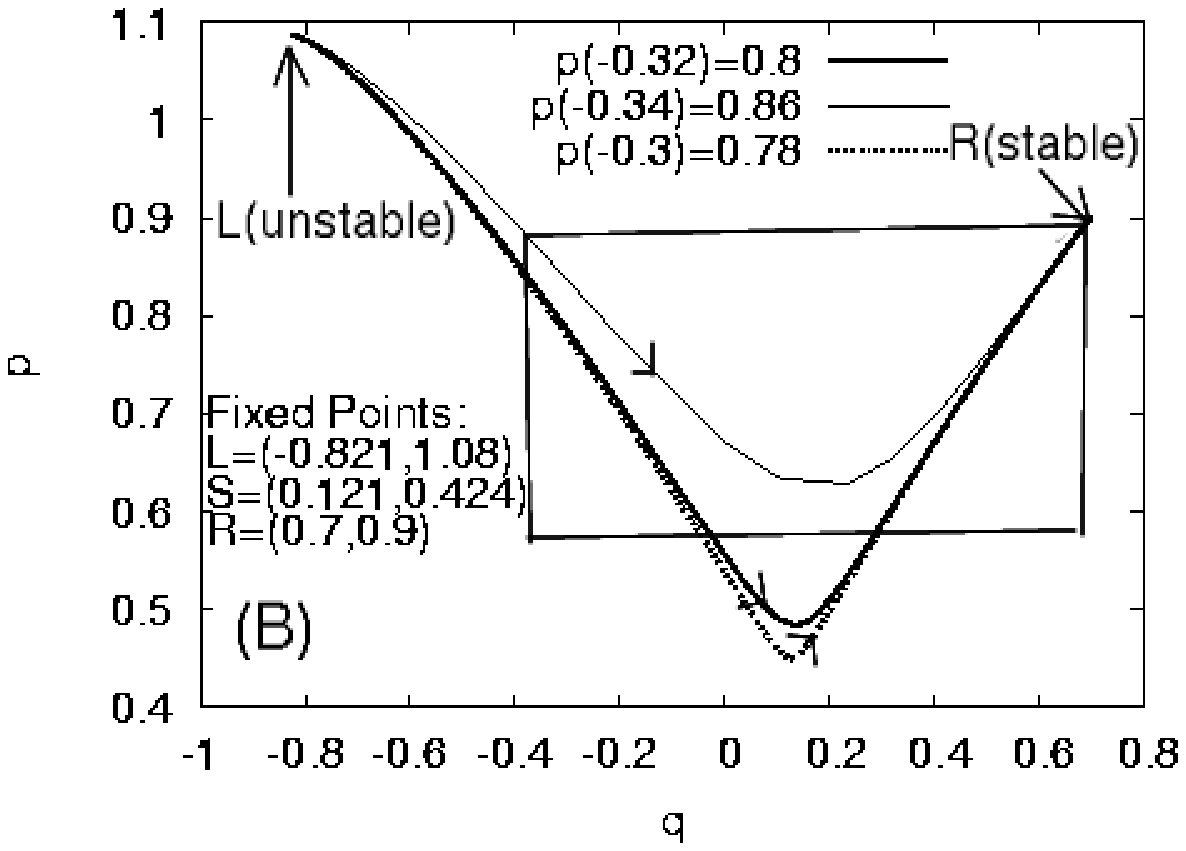}
    \caption{Flow trajectories to the stable fixed point (R)
with the  BCs as specified in figure. Arrows indicate the 
direction of increasing $x$.
 (A) $c=.05$ and $d=-.65$ with BCs specified at $x=1$.
 (B) $c=-.07$ and $d=-.41$ with  BCs at $x=0$.
The rectangular boxes  enclose
 portions of the trajectories with the
 BCs $p(.21)=.95$ for (A) and $p(-.34)=.86$ for (B). 
The length of the box 
is fixed by the fixed point in the physical domain and the 
BC of  the specific  trajectory  it encloses. The vertical 
width of the box is   chosen 
in such a way that it encloses the relevant part of the 
trajectory of interest.  
These parts of the trajectories in A(B) are  seen as  
the right(left)  boundary layers in $\rho$-$x$
and $\eta$-$x$ plane.}
\label{fig:fig_phase1}
\end{figure}

There exist the following possibilities: 

1. L is unstable and lies within the physical domain and R is stable with 
complex conjugate eigenvalues having negative real parts 
(spirally inward flow). R remains outside 
the physical domain and the bulk is, therefore, given by the 
L fixed point values. The bulk in this case is also the same as the left 
BC. The right BC is satisfied by a  boundary layer which is 
a part of the trajectory on the $p-q$ plane 
 starting from L and reaching the stable 
fixed point R. R is outside the 
physical range of $p$ so that any physical BC 
 is satisfied by the right boundary layer. Trajectories 
corresponding to    different  BCs at $x=1$
 are shown in figure 
(\ref{fig:fig_phase1}A). 

2. R is stable and lies within the physical domain and L is unstable with 
complex conjugate eigenvalues (spirally outward flow) having positive 
real parts. 
By the above argument, here the bulk satisfies the right BC
 with  a boundary layer at $x=0$ satisfying the left BC. 
The L fixed point must lie outside the physical domain.
The trajectories in this case  are shown in figure (\ref{fig:fig_phase1}B). 


3. There exists another  situation where  the saddle fixed 
point and  either one of R or L are in the physical domain but the latter 
is different from the BC.
In this case, the  boundary layers should be described by the 
separatrices such that 
the bulk densities can acquire  the  saddle fixed point values. BCs 
at both the boundaries are satisfied by boundary  layers which 
are parts of the separatrices joining L with S, and R with S. This phase 
where the bulk values of $\rho$ and $\eta$
 are given by the saddle fixed point values is the 
intermediate phase.   

A general inspection regarding the stability  
 of the fixed points in  the allowed range of the $c-d$ plane 
confirms these rules as the only possibilities. 
It also  indicates that, for $c>0$ ($c<0$), the
 system remains either in L(R) or in the 
intermediate phase. 
In general, the intermediate phase appears when S is the only  fixed 
point that lies 
in the  domain specified by the BC.

\begin{figure}[htbp]
   \includegraphics[width=3in,clip]{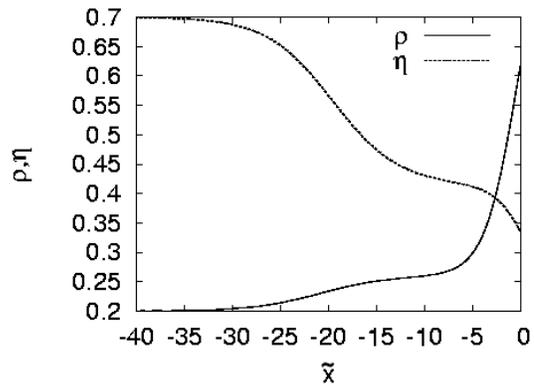}
    \caption{Numerical solutions for the 
boundary layers in the L phase near $x=1$ for $c=.05$ and 
$d=-.65$. The boundary layers of $\rho$ and $\eta$ profile 
 merge to the bulk values $.2$ and $.7$ respectively.  The 
solutions tends to saturate to the saddle fixed point given by 
$(q,p)=(-.153,.673)$}
\label{fig:bl_in_p-x_plane}
\end{figure}

The three rules mentioned above 
completely decide the phases and the phase boundaries. 
Suppose, we are in the left phase where the bulk densities are same 
as the left BCs. This allows us 
to calculate $c$ and $d$, and thereby, the explicit values 
of all the three fixed points. The fixed point corresponding to
 the bulk-density values,
is an unstable one and the boundary layer solutions  approach these 
values as $\tilde x\rightarrow -\infty$  while approaching 
the stable fixed point, which lies beyond the physical region,
 as $\tilde x \rightarrow \infty$. 
 The boundary layers which are part of the trajectories 
in figure (\ref{fig:fig_phase1}),
appear as in figure (\ref{fig:bl_in_p-x_plane}) in the $\rho$-$x$ 
or $\eta$-$x$ plane.
  The trajectories deviate because of the proximity of  
the saddle fixed point and this deviation  appears as 
the tendency of the boundary layer to saturate to an intermediate
value.

\begin{figure}[htbp]
   \includegraphics[width=3in,clip]{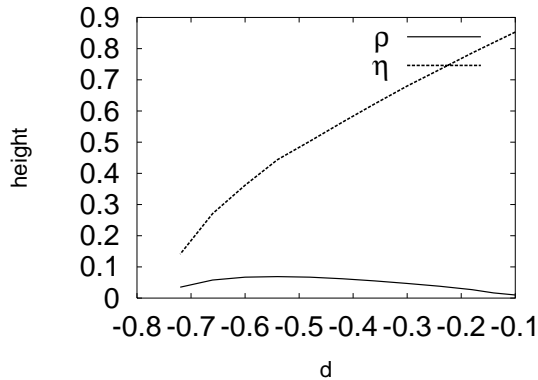}
    \caption{Heights of the newly-formed boundary layers at $x=0$ in the 
 $\rho$ and $\eta$ profiles are plotted  with  $d$ for  
 $c=.05$. These boundary layers are formed at the  phase boundary 
between the L and the intermediate phase.}
\label{fig:heightbl}
\end{figure}

As the right BC is changed further keeping 
the left one unchanged, the solutions still 
appear like those of figure (\ref{fig:fig_phase1}) 
or (\ref{fig:bl_in_p-x_plane})
 with same 
bulk-densities  until a  cusp appears in the trajectory. 
At this right BC, the trajectory is just   
the separatrix joining the unstable and the saddle fixed point on one side 
and  the saddle and the stable fixed point on the other side. Since the 
flow now meets the saddle fixed point, the bulk densities are given by the 
saddle fixed point values. This  requires formation of 
a new boundary layer near 
$x=0$  with the height 
$\rho_u^*-\rho_s^*$ where $\rho_u^*$ and $\rho_s^*$ are the values of 
$\rho$ at the unstable and the saddle fixed points,
 respectively. At this 
point, the system enters the intermediate phase.  
Since the values of the fixed points depend only on 
$c$ and $d$ or on the bulk-density in the L phase, the height of 
the newly formed  boundary layer 
is only dependent on the bulk-density (or on $c$ and $d$)
 of the  L phase. The variation 
of the height of these new boundary layers
with $d$ for a given $c$ is shown in figure (\ref{fig:heightbl}).
    
 The formation of a new boundary layer with a finite 
height and a discontinuous change in the value of the bulk density are 
indications of a first order phase transition.
This basic principle of the bulk densities acquiring the saddle fixed point 
values continues to hold good inside the intermediate phase.
In this  phase, as the right BC is changed, the
bulk values, $p_b$ and $q_b$, change with boundary layers appearing 
at both the boundaries. The  two boundary layers  are the 
two separatrices joining the saddle fixed point to  
the stable and unstable fixed points.  The bulk-densities or the 
 values of $c$ and $d$ are such  that the 
right and left  BCs lie 
 exactly on these two separatrices. 

Next, we  apply 
this method to ASEP with only one kind of particles  moving
rightward  on a 1D lattice 
obeying mutual exclusion. In the steady state, there are 
three phases 
 depending on the left and right boundary densities $\alpha$ and $1-\beta$, 
 \cite{derrida} with the bulk particle-density $\rho_b$ being 
$\rho_b=1-\beta$, for $\alpha>\beta$,  and  $\beta<1/2$, 
$\rho_b=\alpha$,  for $\beta>\alpha$  and $\alpha<1/2$, and  
$\rho_b=1/2$, for  $\alpha>1/2$  and  $\beta>1/2$.
The approach to these phases can be understood well through the motion 
of the domain walls \cite{straley}
 which separate possible steady states of the system.

The equation that describes the boundary layer in this case is 
\begin{eqnarray}
(1/2) \rho'(\tilde x)=\rho(1-\rho)+c,\label{onespbl}
\end{eqnarray}
where the constant $c$ is chosen as $c=-\rho_b(1-\rho_b)$  to assure 
 saturation  of the boundary layer to the bulk as 
$\tilde x\rightarrow \infty\ \ {\rm or}\ \ -\infty$.
This equation has two fixed points $\rho^*=\rho_b, \ 1-\rho_b$. 
A stability analysis around these fixed points shows that 
(i) $\rho_b$ is stable for $\rho_b>1/2$, and 
(ii)$1-\rho_b$ is stable  if $\rho_b<1/2$.

In case of  $\rho_b>1/2$, if a boundary layer has to saturate to 
$\rho_b$ (the stable fixed point),
 its location should be at $x=0$. 
As $\tilde x\rightarrow -\infty$, this boundary layer saturates 
to the unstable 
fixed point $1-\rho_b$.  The right BC is satisfied if 
$\rho_b=1-\beta$.  Two conditions naturally follow from this. 
(i) $1-\beta>1/2$ or $\beta<1/2$ and (ii) $\alpha>1-\rho_b$ or $\alpha>\beta$, 
so that any  boundary layer at $x=0$ satisfies the BC before 
saturating to the fixed point $1-\rho_b$.  
Similarly, in case of $\rho_b<1/2$, the boundary layer saturates to 
$1-\rho_b$  as $\tilde x\rightarrow \infty$ or $\rho_b$  as 
$x\rightarrow -\infty$. A
 boundary layer of this nature should be located at $x=1$  with 
the conditions  (i)$\alpha<1/2$ and (ii) $1-\rho_b>1-\beta$ or
$\alpha<\beta$. As we see here, the location of the boundary layer 
is completely determined by the limit in 
which the bulk fixed point is approached 
by the solution. In both $\rho_b>1/2$ and $\rho_b<1/2$ cases, 
while the boundary layers continue to be in the same position, 
their slopes change across the line $\alpha=1-\beta$. These boundary 
layers correspond to a different set of solutions of equation 
(\ref{onespbl}) and they approach the bulk fixed points at the same 
limits of $\tilde x$ as before with opposite slopes. 
In general, the  boundary layers saturating to two 
different densities are like localized domain walls of reference 
\cite{straley} separating two possible steady states. 
For $\rho_b=1/2$, the two fixed points merge, 
and it is easy to see that the boundary layers from both sides of 
the system  are  able to 
approach the bulk density  in a power law fashion 
as $\sim \frac{1}{\mid\tilde x\mid}$ either as  
$\tilde x\rightarrow  \infty$ or as $\tilde x\rightarrow -\infty$. 
Thus, in this case, there are boundary layers at both the boundaries.      
From a stability-analysis like approach, it is easy to see that  
both the boundary layers  have negative  slopes. 
This implies that such a phase is possible 
when  $\alpha>1/2$ and $\beta>1/2$. These three possibilities 
cover the  entire $\alpha$-$\beta$ parameter space.

In this paper, we show how a phase-plane analysis of the 
differential equations for the boundary layers can be used   
to predict boundary induced phase transitions in ASEP. 
This method is applied to single species as well as two species
processes. 
We arrive at a 
general prediction that for flat bulk profiles, the bulk densities 
 in different   phases 
are given by various fixed point values of the differential 
equations describing the boundary layers. 
For two species case, we  find out 
the bulk density of the intermediate phase which has no one species 
analogue.  The fixed point analysis 
determines the nature of the transitions, 
the location of the boundary layers and 
how the boundary layers merge to the bulk densities. For non-constant 
bulk density, the fixed point of the boundary layer should match 
with the bulk-density value near their merging. 
In addition, we believe that this method can be applied to more 
complicated systems, such as those with more than two species of 
particles and the number of phase transitions can be predicted 
from the locations  of the  fixed points on the phase-plane.  
The generality of the method suggests  its 
wider  applicability  to various other systems where boundary induced 
transitions can be seen. 

Financial support from DST (India) is gratefully acknowledged.


\begin{thebibliography}{99}
\bibitem{comment1} Although 
the spontaneous symmetry breaking is found in these models, there are 
differences  regarding the number and nature  of phase transitions 
obtained from numerical and mean-field analysis \cite{mukamel,arndt}.
\end{thebibliography}
\end{document}